\documentclass[aip,sd,amsmath,amssymb,reprint]{revtex4-1}

\usepackage{graphicx}
\usepackage{dcolumn}
\usepackage{bm}
\usepackage{color}
\usepackage{adjustbox}

\begin{document}

\title{Discrete dynamical systems with scaling and inversion symmetries}

\author{Vaguiner Rodrigues dos Santos}
\email{vaguiner@usp.br} 
\affiliation{Institute of Physics, University of S\~ao Paulo, 05508-090, S\~ao Paulo, SP, Brazil.}
\author{Enrique Chipicoski Gabrick}
\email{ecgabrick@gmail.com}
\affiliation{Institute of Physics, University of S\~ao Paulo, 05508-090, S\~ao Paulo, SP, Brazil.}
\author{Edson Denis Leonel}
\affiliation{Institute of Geosciences and Exact Sciences, S\~ao Paulo State University (UNESP), 13506-900, Rio Claro, SP, Brazil.}
\author{Iber\^e Luiz Caldas}
\affiliation{Institute of Physics, University of S\~ao Paulo, 05508-090, S\~ao Paulo, SP, Brazil.}
%

\begin{abstract}
In this work, we investigate scale invariance in the temporal evolution and chaotic regime of discrete dynamical systems. By exploiting the close interrelation between scaling and inversion transformations, we formulate scale symmetry in terms of inversion symmetry. As applications of our approach, we determine fractal dimensions and compute Lyapunov exponents for paradigmatic dynamical systems using scaling and inversion symmetries. By comparing our method with standard approaches, we obtain identical numerical values for the Lyapunov exponents using only a small number of iterations. Furthermore, our geometric-based framework naturally provides access to the fractal dimension. The agreement with standard results demonstrates that the proposed method is efficient and can be effectively employed in the study of dynamical systems.
\end{abstract}

\maketitle

\begin{quotation}
Several complex systems exhibit scale invariance, i.e., their properties remain unchanged when observed at different scales. In this work, we show that this fundamental feature, particularly in discrete dynamical systems, can be understood through inversion symmetry. By formulating scale invariance in terms of inversion, we propose a unified framework to address two central problems in nonlinear science: the characterization of fractal measures and the quantification of chaos. In particular, we demonstrate that both quantities can be obtained directly from inversion-based scaling relations.
\end{quotation}

\section{Introduction}\label{sec_intro}
Scale invariance is present in a wide variety of dynamical and complex systems, typically manifesting itself through power-law behavior and the emergence of scaling laws \cite{Gl:04}. Prominent examples include fractal systems \cite{Ma:83}, critical phenomena \cite{An:12}, and other nonlinear dynamical systems \cite{Le:19}.

Scale invariance has been extensively investigated across different fields, particularly in classical field theory \cite{add1}, statistical physics \cite{add2}, and chaos theory \cite{add3}, as well as in classical electromagnetism \cite{Ba:10}. In his seminal work, Bateman (see Ref.~\cite{Ba:10}) identified the full set of transformations under which the classical electrodynamics equations are invariant, revealing conformal invariance and, as a special case, scale invariance of Maxwell's equations. This result extended Lorentz symmetry in Minkowski spacetime. From the second half of the twentieth century onward, scale invariance became a central concept in statistical physics \cite{add4,add5}, particularly in the study of phase transitions and critical phenomena, most notably through the work of Kenneth G.~Wilson on the renormalization group \cite{Wi:75}. In the context of chaotic systems, the Huberman--Rudnick scaling law \cite{Hu:80} provides an example of how scaling behavior is directly related to the emergence of positive Lyapunov exponents near the transition to chaos.

A scale-invariant system remains unchanged under dilatations or contractions by a constant scale factor. Scale transformations themselves can be constructed as compositions of inversion transformations \cite{Bl:00, Co:67}, establishing a direct connection between scaling and inversion symmetries.

Inversion is a discrete transformation, essentially defined as a reflection with respect to a circle. It can be formulated in low dimensions, on hyperspheres, or extended to higher-dimensional spaces where pairs of points are transformed into one another. Beyond its fundamental role in solving geometric problems in several areas \cite{Ma:07}, inversion is widely used in conformal mapping techniques \cite{Ne:97}. Notably, inversion symmetry is intrinsically present in Maxwell's equations \cite{Ka:08} and plays a central role in conformal field theories \cite{Ka:08, Bl:09}.

The first application of inversion transformations in physics is commonly attributed to Lord Kelvin. In 1845, Kelvin introduced and applied inversion techniques to solve electrostatic problems involving charged spheres \cite{Th:45, Th:72}. Later, in 1910, Ebenezer Cunningham, together with Bateman, extended the laws of electromagnetism to include inversion transformations \cite{Cu:10}. As with scale transformations, inversion symmetry is also fundamental in two-dimensional conformal field theories \cite{Bl:09}, which are characterized by an infinite number of local symmetries.

Given the intimate relationship between scale and inversion symmetries, in this work we propose a formulation of scale invariance explicitly in terms of inversion transformations. We illustrate our approach through two main applications. The first concerns the determination of fractal dimensions in self-similar systems \cite{Ma:83}, while the second focuses on the estimation of Lyapunov exponents for paradigmatic dynamical maps \cite{Al:96}. Whereas the inversion symmetry of self-similar fractals provides a natural framework for computing their fractal dimension, the evaluation of Lyapunov exponents based on symmetry considerations offers a simplified and efficient alternative to standard numerical procedures.

This paper is organized as follows. In Section~\ref{sec2}, we introduce the geometric inversion transformation and discuss its relationship with scaling transformations. In Section~\ref{sec3}, we present a set of definitions relevant to systems exhibiting inversion symmetry. In Section~\ref{sec_4}, we apply our methodology to self-similar fractals. Section~\ref{sec_5} is devoted to the computation of Lyapunov exponents using scale and inversion symmetries. Finally, in Section~\ref{sec_conclusion}, we summarize our results and draw concluding remarks.

\section{Scaling transforms and inversion}\label{sec2}
A scaling transformation $S(x)$ corresponds to an expansion or contraction along a given direction $x$ and is defined by
\begin{equation}
S(x)=sx, \label {S(x)} 
\end{equation}
where the scale factor $s>0$ is a real, non-unitary constant \cite{Co:67}. According to this definition, a contraction occurs when $0<s<1$, whereas an expansion corresponds to $s>1$.

In this work, we restrict the inversion transformation to the plane 
\cite{Co:67}, commonly referred to as geometric inversion, defined with 
respect to a circle, hereafter called the inversion circumference.

\begin{figure}[hbt]
\centering
\includegraphics[scale=0.45]{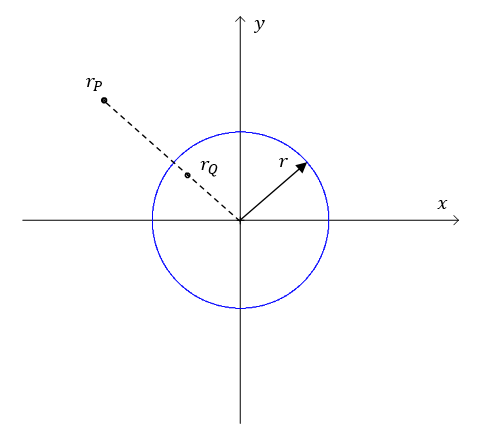}
\caption{The inversion transformation in the plane, defined by $r_P.r_Q=r^2$, which maps external points $P$ into internal points $Q$ and vice versa.}
\label{fig1}
\end{figure}

According to Fig.~\ref{fig1}, plane inversion relates two points $P(x_P,y_P)$ and $Q(x_Q,y_Q)$ with respect to a circumference of radius $r$ and center $O(x_0,y_0)$. Under inversion, a point $P$ located at a distance $r_P\geq r$ from the center is mapped onto a point $Q$ located at a distance $r_Q\leq r$. The inversion process is governed by the relation
\begin{equation}                                           
r_P.r_Q=r^2, \label {rPQ}   
\end{equation}
where $r_P=\sqrt{(x_P-x_0)^2+(y_P-y_0)^2}$ and $r_Q=\sqrt{(x_Q-x_0)^2+(y_Q-y_0)^2}$.

The scale transformation in Eq.~(\ref{S(x)}) can be constructed as a composition of two consecutive inversion transformations. Considering the inversion maps $T(x)=\frac{r^2}{x}$ and $T'(x)=\frac{r'^2}{x}$ associated with circumferences of radii $r$ and $r'$, respectively, one obtains $S(x)=T'(x)\circ T(x)$, with the corresponding scale factor given by $s=\left(\frac{r'}{r}\right)^2$.

Scale and inversion transformations belong to the group of conformal transformations, and the symmetry associated with this group is referred to as conformal symmetry \cite{Ka:08,Bl:09}.
\section{Definitions: inverse sets, functions, and inversion results}\label{sec3}
We now introduce the key concepts required for the development of our methodology, including inverse sets, inversion functions, and inversion resultants. Based on these definitions, we derive the differential equations that characterize systems exhibiting inversion symmetry.

\subsection{Inverse sets}
The inversion transformation along the radial direction, given by Eq.~(\ref{rPQ}), relates the set of points $I_{1}=\left]0,r\right]$ to the set $I_{2}=\left[r,+\infty\right[$ with respect to a circumference of radius $r$. We refer to $I_{1}$ and $I_{2}$ as inverse sets.

Motivated by Eq.~(\ref{rPQ}), we propose the following relation between elements $x_{j}$ and $x_{\ell}$ of one-dimensional systems exhibiting inversion symmetry, defined with respect to a parameter $k_{x}$:
\begin{equation}
x_{j}.x_{\ell}=k_{x}, \label{relacao de inversao kx} 
\end{equation}
with $x_{j}$, $x_{\ell}$, $k_{x}\in \mathbb{R}^{*}$ and $j,\ell=1,2$. Equation~(\ref{relacao de inversao kx}) can be generalized to higher dimensions as well as to the complex domain.

The inverse sets associated with Eq.~(\ref{relacao de inversao kx}) are defined in terms of the absolute value of $k_{x}$. On the real $x$-axis, we define four inverse sets:
\begin{eqnarray} 
I^{+}_{1x}&=&\left]0,+\sqrt{|k_{x}|}\right], \\
I^{-}_{1x}&=&\left[-\sqrt{|k_{x}|},0\right[, \\
I^{+}_{2x}&=&\left[+\sqrt{|k_{x}|},+\infty\right[, \\ 
I^{-}_{2x}&=&\left]-\infty,-\sqrt{|k_{x}|}\right].
\end{eqnarray} 

For $k_{x}>0$, the sets $I^{+}_{1x}$ and $I^{+}_{2x}$ are symmetric inverses of one another, as are the sets $I^{-}_{1x}$ and $I^{-}_{2x}$. For $k_{x}<0$, the symmetric inverse pairs are $I^{+}_{1x}$ with $I^{-}_{2x}$, and $I^{-}_{1x}$ with $I^{+}_{2x}$. The union of all four inverse sets defines the set $I_{x}=I^{+}_{1x}\cup I^{-}_{1x}\cup I^{+}_{2x}\cup I^{-}_{2x}$, which explicitly excludes the point $x=0$.

We say that a one-dimensional discrete dynamical system exhibits inversion symmetry when two orbits, denoted by $O_{x}$ and $O'_{x}$, belong to inverse sets. In this case, the elements $x$ of orbit $O_{x}$ and the elements $x'$ of orbit $O'_{x}$ are related through Eq.~(\ref{relacao de inversao kx}), leading to
\begin{equation}
x.x'=\pm (x^{*})^2,
\end{equation}
where $x^{*}=\pm \sqrt{|k_{x}|}$ is a non-null fixed point of the system.

\begin{figure}[hbt]
\centering
\includegraphics[scale=0.45]{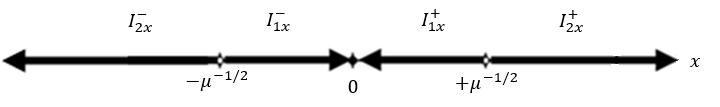}
\caption{Phase portrait of the map $x_{m+1}=\mu x_{m}^3$ for $\mu>0$, where the orbit generated from a non-null initial condition $x_0$ belongs to one of the inverse sets $I^{+}_{1x}$, $I^{+}_{2x}$, $I^{-}_{1x}$, or $I^{-}_{2x}$.}
\label{fig2}
\end{figure}

The phase portrait of the map $x_{m+1}=\mu x_{m}^3$ for $\mu>0$ is shown in Fig.~\ref{fig2}. This map possesses three fixed points, namely $x_{1}^{*}=0$, $x_{2}^{*}=\mu^{-1/2}$, and $x_{3}^{*}=-\mu^{-1/2}$. For a non-null initial condition $x_0$, the corresponding orbit belongs to one of the inverse sets $I^{+}_{1x}$, $I^{+}_{2x}$, $I^{-}_{1x}$, or $I^{-}_{2x}$, associated with the inversion relations $x.x'=-\mu^{-1}$ and $x.x'=+\mu^{-1}$.

\subsection{Resulting from inversion}
The sequences of elements belonging to the inverse sets defined by Eq.~(\ref{relacao de inversao kx}) can be naturally represented in vector form.

While the sequences associated with the sets $I^{+}_{2x}$ and $I^{-}_{2x}$ diverge to $+\infty$ and $-\infty$, respectively, the sequences corresponding to the sets $I^{+}_{1x}$ and $I^{-}_{1x}$ converge toward zero.

These vectorial trends are clearly reflected in the orbits of one-dimensional discrete dynamical systems exhibiting inversion symmetry, as illustrated by the phase portrait shown in Fig.~\ref{fig2}.

We define the inversion resultant associated with elements $x_{j}$ and $x_{\ell}$ of the inverse sets given by Eq.~(\ref{relacao de inversao kx}) as
\begin{equation}
\vec R_{j,\ell}(x)=\left(x_{j}-x_{\ell}\right)\vec e_{x}, \label {resultantejl} 
\end{equation}
with $j,\ell=1,2$.

\subsection{Inversion characteristic function and differential equation}
Considering Eq.~\eqref{resultantejl} for the inverse elements $x_{j}$ and $x_{\ell}$, we obtain the following equation for each element as a function of $R_{j,\ell}(x)=|\vec R_{j,\ell}(x)|$ and the parameter $k_{x}$:
\begin{equation}
x_{j,\ell}^{2}\pm x_{j,\ell} R_{j\ell}(x)-k_{x}=0.
\end{equation}

The inverse elements can therefore be expressed as functions that depend on the parameter $k_{x}$ and on the resultant $R_{j\ell}(x)$ associated with these elements, namely
\begin{equation}
x_{j,\ell}=\mp \frac{R_{j\ell}(x)}{2}\pm\sqrt{\left[\frac{R_{j\ell}(x)}{2}\right]^{2}+k_{x}}. \label {função kR} 
\end{equation}

From Eq.~\eqref{função kR}, we infer the existence of a characteristic inversion function, generically defined in terms of a constant parameter $k\equiv k_{x}$ and a variable $\varepsilon\equiv \frac{R_{j\ell}(x)}{2}$, which is related to the resultant vector associated with the inverse elements. This function is defined as
\begin{equation}
f_{i}^{(\alpha,k)}(\varepsilon)=-\varepsilon-(-1)^{\alpha}h_{k}(\varepsilon), \label {definição de funçao de inversao} 
\end{equation}
where $\alpha=1,2$ labels the inversion branch, $f_{i}^{(1,k)}(\varepsilon).f_{i}^{(2,k)}(\varepsilon)=-k$, and $h_{k}(\varepsilon)=\sqrt{\varepsilon^2+k}$. The function $f_{i}^{(\alpha,k)}(\varepsilon)$ determines the inverse elements of systems exhibiting inversion symmetry.

Differentiating both sides of Eq.~\eqref{definição de funçao de inversao} with respect to $\varepsilon$, we obtain the differential equation associated with inversion-symmetric systems, given by
\begin{equation}
\frac{df_{i}^{(\alpha,k)}(\varepsilon)}{d\varepsilon}=(-1)^{\alpha}\frac{f_{i}^{(\alpha,k)}(\varepsilon)}{h_{k}(\varepsilon)}. \label {equação diferencial de inversao} 
\end{equation}

From Eq.~\eqref{equação diferencial de inversao}, the quantities $\varepsilon$ and $h_{k}(\varepsilon)$ can be expressed in terms of the inversion functions for a given system with inversion symmetry. These relations follow from
\begin{eqnarray}
&&f_{i}^{(1,k)}(\varepsilon).f_{i}^{(2,k)}(\varepsilon)=-k, \label {inversao Ik} \\ 
\varepsilon&=&\frac{1}{2}\left[\frac{k}{f_{i}^{(\alpha,k)}(\varepsilon)}-f_{i}^{(\alpha,k)}(\varepsilon)\right], \label {epsilon} \\
h_{k}(\varepsilon)&=&\frac{1}{2}(-1)^{\alpha+1}\left[\frac{k}{f_{i}^{(\alpha,k)}(\varepsilon)}+f_{i}^{(\alpha,k)}(\varepsilon)\right]. \label {hk} 
\end{eqnarray}

\subsection{Characteristic function of inversion in exponential form}
The solution of Eq.~\eqref{equação diferencial de inversao} for the inversion function $f_{i}^{(\alpha,k)}(\varepsilon)$ can be written as
\begin{equation}
f_{i}^{(\alpha,k)}(\varepsilon)=a_{\alpha}e^{\varphi^{(\alpha,k)}(\varepsilon)}, \label {solução da equação diferencial}
\end{equation}
where $\varphi^{(\alpha,k)}(\varepsilon)=\int\frac{(-1)^{\alpha}d\varepsilon}{h_{k}(\varepsilon)}$, with $\alpha=1,2$, and where the constants satisfy $f_{i}^{(1,k)}(\varepsilon).f_{i}^{(2,k)}(\varepsilon)=a_{1}.a_{2}=-k$.

Let us now consider the following parametrization for $k$ and $\varepsilon$:
\begin{eqnarray}
k&=&1, \\
\varepsilon(\gamma)&=&\sinh(\gamma)=\frac{1}{2}\left(e^{\gamma}-e^{-\gamma}\right).
\end{eqnarray}
Under this parametrization, we obtain
\begin{equation}
\varphi^{(\alpha,1)}(\gamma)=\int (-1)^{\alpha}d\gamma=(-1)^{\alpha}\gamma.
\end{equation}

The inversion function in Eq.~\eqref{solução da equação diferencial} can then be written in the exponential form
\begin{equation}
f_{i}^{(\alpha)}(\gamma)=a_{\alpha}e^{(-1)^{\alpha}\gamma}, \label {função de inversão com expoente} 
\end{equation}
with $\alpha=1,2$. Finally, we define $\gamma$ as the inversion exponent. It is worth emphasizing that when $k=1$ the inversion function is normalized, allowing us to omit the parameter $k$ from the notation, such that $f_{i}^{(\alpha,k)}(\varepsilon)=f_{i}^{(\alpha,1)}(\varepsilon)=f_{i}^{(\alpha)}(\varepsilon)$.

\section{Scale and inversion symmetries in fractals}\label{sec_4}
Fractals are well known for their inherent scale symmetry \cite{Ma:83}. In this section, we explore how measurements of length, area, and volume in fractal structures can be represented in terms of inverse sets, and how the fractal dimension can be determined through inversion symmetry.

\subsection{Geometric inverse sets}
We begin by considering inverse sets formed from discrete measurements of the perimeter, area, or volume of a given geometric figure.

As an illustrative example, let us consider a geometric figure initiated from an equilateral triangle $\Delta_{0}$ with side length $l_0$. The corresponding geometric measures are the perimeter, given by $P_{0}=3l_{0}$, and the area, given by $S_0=\ell_0^2\frac{\sqrt{3}}{4}$.

For the perimeter $P_n$ of triangles $\Delta$ belonging to the inverse set $I_{2}^{+}$, which increase according to a scale factor $\rho>1$, the corresponding measures $P'_{n}$ of triangles $\Delta'$ belonging to the inverse set $I_{1}^{+}$ decrease according to the contraction factor $c=1/\rho$. These relations are expressed as
\begin{eqnarray}
P_{n}&=&\rho^{n} P_0, \label {perímetro_fi1} \\
P'_{n}&=&c^{n}P_0=\rho^{-n} P'_0, \label {perímetro_fi2} 
\end{eqnarray}
where $n=0,1,2,\ldots$ and $P'_0=P_0$.

Accordingly, the inverse sets $I_{1}^{+}$ and $I_{2}^{+}$ can be written as
\begin{eqnarray}
I_{1}^{+}&=&\left]0,\ldots,\frac{P_0}{\rho^3},\frac{P_0}{\rho^2},\frac{P_0}{\rho},P_0\right], \\
I_{2}^{+}&=&\left[P_0,\rho P_0,\rho^2 P_0,\rho^3 P_0,\ldots\right[.
\end{eqnarray}

\begin{figure}[hbt]
\centering
\includegraphics[scale=0.45]{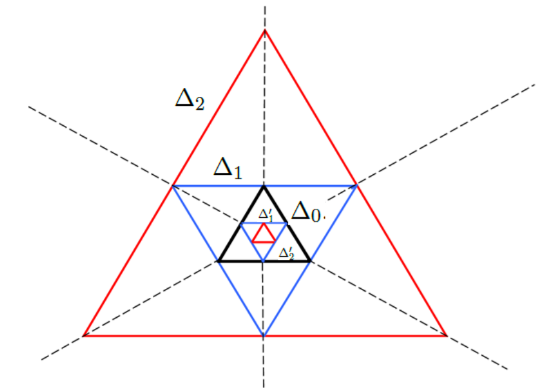}
\caption{A geometric figure resulting from the union of the inverse sets formed by equilateral triangles.}
\label{fig3}
\end{figure}

The geometric figure obtained from the union of these two inverse sets is shown in Fig.~\ref{fig3}. In this construction, the perimeter measures of triangles $\Delta_1$ and $\Delta_2$ are inversely related to the corresponding measures of $\Delta_1'$ and $\Delta_2'$ with respect to the reference triangle $\Delta_0$, namely,
\begin{equation}
P'_{n}.P_{n}=P_0^{2}. \label {perímetros inversos} 
\end{equation}

Normalizing Eq.~\eqref{perímetros inversos} as $\left(\frac{P'_n}{P_0}\right)\left(\frac{P_n}{P_0}\right)=1$, we define the normalized inversion functions as
\begin{eqnarray}
f_{i}^{(1)}(n)&=&\frac{P'_n}{P_0}=\rho^{-n}, \label {normalizadan1} \\
f_{i}^{(2)}(n)&=&-\frac{P_n}{P_0}=-\rho^{n}, \label {normalizadan2} 
\end{eqnarray}
which satisfy $f_{i}^{(1)}(n).f_{i}^{(2)}(n)=-1$.

Using Eqs.~\eqref{inversao Ik}, \eqref{epsilon}, and \eqref{hk}, the normalized inversion functions defined in Eqs.~\eqref{normalizadan1} and \eqref{normalizadan2} are solutions of the inversion differential equation \eqref{equação diferencial de inversao} for the perimeters shown in Fig.~\ref{fig3}, yielding
\begin{equation}
\frac{df_{i_P}^{(\alpha)}(n)}{dn}=(-1)^{\alpha+1}{\rm ln}(c)f_{i_P}^{(\alpha)}(n). \label {diferencial_perímetro}
\end{equation}

The areas associated with the inverse sets $I_{1}^{+}$ and $I_{2}^{+}$ are given, respectively, by $S_{n}=\rho^{2n}S_0$ and $S'_{n}=\rho^{-2n}S_0$.

The corresponding normalized inversion functions for the areas are
\begin{eqnarray}
f_{i_S}^{(1)}(n)&=&\frac{S'_n}{S_0}=\rho^{-2n}, \\
f_{i_S}^{(2)}(n)&=&-\frac{S_n}{S_0}=-\rho^{2n}.
\end{eqnarray}

The inversion differential equation associated with the area measurements then takes the form
\begin{equation}
\frac{df_{i_S}^{(\alpha)}(n)}{dn}=2(-1)^{\alpha+1}{\rm ln}(c)f_{i_S}^{(\alpha)}(n). \label {diferencial_área}
\end{equation}

Equations~\eqref{diferencial_perímetro} and \eqref{diferencial_área} have the same functional structure. The factor $d_E=2$ appearing in Eq.~\eqref{diferencial_área} is associated with the type of geometric measure. In Eq.~\eqref{diferencial_perímetro}, the perimeter corresponds to a one-dimensional measure with Euclidean dimension $d_E=1$, whereas in Eq.~\eqref{diferencial_área} the area corresponds to a two-dimensional measure with $d_E=2$.

More generally, the inversion differential equation associated with the geometric measures of a given figure depends on the linear contraction factor $c$ and on the Euclidean dimension $d_E$, and can be written as
\begin{equation}
\frac{df_{i_{P,S,V}}^{(\alpha)}(n)}{dn}=(-1)^{\alpha+1}{\rm ln}(c)f_{i_{P,S,V}}^{(\alpha)}(n)d_E, \label {equação direrencial fgi} 
\end{equation}
where $f_{i_{P,S,V}}^{(\alpha)}(n)$ denotes the normalized inversion function associated with the perimeter (P), area (S), or volume (V) of the geometric figure, and $d_E$ is the Euclidean dimension of the corresponding measure. Linear measures have $d_E=1$, surface measures have $d_E=2$, and volume measures have $d_E=3$.
 
\subsection{Inverse sets and inversion differential equation for self-similar fractals}
The length, area, and volume measurements of fractals are also naturally related to geometric inverse sets. To investigate the inversion symmetry associated with fractals, we focus on self-similar fractals, which exhibit scale invariance throughout their structure.

Self-similar fractals are characterized by the following power-law relation:
\begin{equation}
\tau(\kappa)\propto\kappa^{-d_F}, \label {lei de potencia} 
\end{equation}
where $\tau(\kappa)$ denotes the minimum number of $N$-dimensional hypercubes of side length $\kappa$ required to cover the entire set of points in an $N$-dimensional space, and $d_F$ is the fractal dimension.

As a concrete example, we analyze the perimeter and area of the Sierpinski triangle \cite{Ma:83} and examine the inversion symmetry associated with this fractal structure.

\begin{figure}[hbt]
\centering
\includegraphics[scale=0.45]{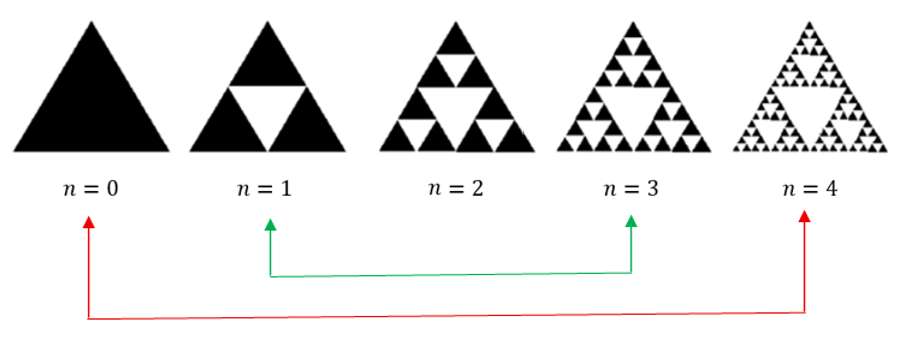}
\caption{Sierpinski triangle up to iteration $n=4$, illustrating that the perimeter and area measurements are inversely related with respect to iteration $n=2$.}
\label{fig4}
\end{figure}

Figure~\ref{fig4} illustrates the inversion symmetry observed in the Sierpinski triangle. In this construction, the iteration measures corresponding to $n=1$ and $n=3$ form an inverse pair with respect to the iteration $n=2$, while the iteration measures $n=0$ and $n=4$ constitute another inverse pair.

The perimeter at each iteration is governed by the recurrence relation
\begin{equation}
P_n=\left(\frac{3}{2}\right)^n P_0, \label {perímetro TS} 
\end{equation}
where $P_0=3\ell_0$ denotes the perimeter at iteration $n=0$.

Since the fractal is not defined for $n<0$, the perimeter measurements are associated exclusively with the inverse set $I_{1}^{+}$, which can be written as
\begin{equation}
I_{1}^{+}=\left]0,\ldots,\frac{27P_0}{8},\frac{9P_0}{4},\frac{3P_0}{2},P_0\right].
\end{equation}

The perimeters belonging to the inverse set $I_{1}^{+}$ satisfy the following inversion relation:
\begin{equation}
P_{(n-q)}.P_{(n+q)}=(P_n)^2, \label {inversaoxq} 
\end{equation}
where $n\geq q$ and $q=1,2,\ldots,n$.

For instance, with respect to $n=1$, the inverse perimeters are $P_0$ and $P_2$ for $q=1$. Similarly, for $n=2$, the inverse pairs are $P_1$ and $P_3$ for $q=1$, and $P_0$ and $P_4$ for $q=2$, and so forth.

The normalized inversion functions corresponding to the perimeter are therefore given by
\begin{eqnarray}
f_{i_P}^{(1)}(q)&=&\frac{P_{(n-q)}}{P_n}=\left(\frac{3}{2}\right)^q, \\
f_{i_P}^{(2)}(q)&=&-\frac{P_{(n+q)}}{P_n}=-\left(\frac{3}{2}\right)^{-q}.
\end{eqnarray}

The differential equation governing the inversion symmetry of the Sierpinski triangle then reads
\begin{equation}
\frac{df_{i_{P}}^{(\alpha)}(q)}{dq}=(-1)^{\alpha}{\rm ln}\left(\frac{3}{2}\right)f_{i_{P}}^{(\alpha)}(q). \label {diferencial TS} 
\end{equation}

Equation~\eqref{diferencial TS} can be rewritten by introducing the fractal dimension $d_F=\frac{{\rm ln}(3)}{{\rm ln}(2)}$ and the contraction factor $c=\frac{1}{2}$ of the Sierpinski triangle, yielding
\begin{equation}
\frac{df_{i_{P}}^{(\alpha)}(q)}{dq}=(-1)^{\alpha+1}f_{i_{P}}^{(\alpha)}(q){\rm ln}(c)(d_F-1).
\end{equation}

The area at each iteration follows the recurrence relation $S_n=\left(\frac{3}{4}\right)^n S_0$, where $S_0=\ell_0^2\frac{\sqrt{3}}{4}$ is the area at iteration $n=0$.

The area measurements are likewise associated with the inverse set $I_{1}^{+}$, given by
\begin{equation}
I_{1}^{+}=\left]0,\ldots,\frac{27S_0}{48},\frac{9S_0}{16},\frac{3S_0}{4},S_0\right].
\end{equation}

The inversion differential equation associated with the area measurements of the Sierpinski triangle is
\begin{equation}
\frac{df_{i_{S}}^{(\alpha)}(q)}{dq}=(-1)^{\alpha+1}f_{i_{S}}^{(\alpha)}(q){\rm ln}(c)(d_F-2).
\end{equation}

More generally, for other deterministic self-similar fractals, one finds a characteristic inversion differential equation analogous to Eq.~\eqref{equação direrencial fgi}, which depends on the contraction factor $c$, the fractal dimension $d_F$, and the Euclidean dimension $d_E$, and can be written as
\begin{equation}
\frac{df_{i_{P,S,V}}^{(\alpha)}(q)}{dq}=(-1)^{\alpha+1}f_{i_{P,S,V}}^{(\alpha)}(q){\rm ln}(c)\delta d,
\end{equation}
where $\delta d=d_F-d_E$, and $f_{i_{P,S,V}}^{(\alpha)}(q)$ denotes the normalized inversion function associated with the perimeter (P), area (S), or volume (V) of the fractal.

In this case, the corresponding solutions for the first and second inversion functions are
\begin{eqnarray}
f_{i_{P,S,V}}^{(1)}(q)&=&c^{q(d_F-d_E)}, \label {primeira função}  \\
f_{i_{P,S,V}}^{(2)}(q)&=&-c^{-q(d_F-d_E)}, \label {segunda função} 
\end{eqnarray}
which satisfy $f_{i_{P,S,V}}^{(1)}(q).f_{i_{P,S,V}}^{(2)}(q)=-1$.

Therefore, the scale symmetry of fractals is described by the power law given in Eq.~\eqref{lei de potencia}, whereas the inversion symmetry manifests itself through exponential laws, namely Eqs.~\eqref{primeira função} and \eqref{segunda função}.

\section{Scale and inversion symmetries in chaotic maps}\label{sec_5}
In this section, we estimate the Lyapunov exponents of one-dimensional chaotic maps by exploiting scale and inversion symmetries.

\subsection{Lyapunov exponents using scale and inversion symmetries}
Lyapunov exponents are commonly used to determine whether a dynamical system exhibits chaotic behavior \cite{Al:96,Ca:17}. To introduce them, consider a one-dimensional map defined by $x_{n+1}=F(x_n)$. For two orbits generated from nearby initial conditions $x$ and $x'$, the Lyapunov exponent $\lambda$ is defined such that the distance between the orbits evolves exponentially, namely,
\begin{equation}
\delta F^m\approx\delta_x e^{m\lambda}, \label {diferencial delta Fm} 
\end{equation}
where $F^m(x)$ denotes the mapping of order $m$, corresponding to the $m$-th iteration of the function $F(x)$.

The Lyapunov exponent converges in the limit $m\rightarrow\infty$, yielding
\begin{equation}
\lambda=\lim_{m\rightarrow\infty}\frac{1}{m}\sum_{j=0}^{m-1}{\rm ln}\left|F'(x_j)\right|. \label {expoente de Lyapunov} 
\end{equation}

As will be shown below, one-dimensional discrete dynamical systems may exhibit scale and inversion symmetries in the asymptotic limit $m\rightarrow\infty$. By exploiting these symmetries, we establish a direct relationship between the inversion exponent and the Lyapunov exponent of one-dimensional chaotic maps.

Let $F^m(x)$ denote the mapping of order $m$ of a one-dimensional discrete dynamical system. Consider the curve associated with the mapping $F^m(x)$ defined over an interval $\Delta x=x_b-x_a$. The length $L$ of a curve $f(x)$, provided that the derivative $\frac{df(x)}{dx}$ exists in the interval $\Delta x$, is given by the standard expression \cite{Ka:72}
\begin{equation}
L=\int_{x_a}^{x_b}\sqrt{1+\left|\frac{df(x)}{dx}\right|^2}\,dx.
\end{equation}

Accordingly, the length $L(m)$ of the curve associated with the mapping $F^m(x)$ is
\begin{equation}
L(m)=\int_{x_a}^{x_b}\sqrt{1+\left|\frac{dF^m(x)}{dx}\right|^2}\,dx. \label {L(m) integral} 
\end{equation}

To estimate $L(m)$ numerically over the interval $\Delta x$, we discretize the curve into $K$ segments, leading to
\begin{equation}
L(m)\approx \sum_{i=1}^{K}\sqrt{\Delta x_i^2+\Delta F^m(x_i)^2},
\end{equation}
where $\Delta x_i=x_i-x_{i-1}$ and $\Delta F^m(x_i)=F^m(x_i)-F^m(x_{i-1})$. The smaller the interval $\sigma_x=\Delta x/K$ between consecutive segments, the more accurate the approximation of the curve length.

Considering Eq.~\eqref{diferencial delta Fm} in the asymptotic regime $\delta_x\rightarrow 0$, we obtain
\begin{equation}
\left|\frac{dF^m(x)}{dx}\right|\approx e^{m\lambda}. \label {Lyapunov assintotico} 
\end{equation}

If the derivative $\frac{dF^m(x)}{dx}$ exists over the interval $\Delta x$, substitution of Eq.~\eqref{Lyapunov assintotico} into Eq.~\eqref{L(m) integral} yields
\begin{equation}
L(m)\approx\int_{x_a}^{x_b}\sqrt{1+\left(e^{m\lambda}\right)^2}\,dx. \label {comprimento}  
\end{equation}

A defining characteristic of chaotic systems is $\lambda>0$. Since $m>0$, in the asymptotic regime $m\lambda\rightarrow\infty$ we have $e^{m\lambda}\gg1$, which leads to
\begin{equation}
L(m)\approx e^{m\lambda}\Delta x. \label {comprimento2}  
\end{equation}

Therefore, in the asymptotic limit, the lengths of the curves associated with the mappings of order $m$ and $m+1$ satisfy
\begin{equation}
L(m+1)\approx e^{\lambda}L(m). \label {comprimento3} 
\end{equation}

Equation~\eqref{comprimento3} has the form $x_{m+1}=\rho x_m$, where the scale factor $\rho=e^{\lambda}$ characterizes the scale symmetry of the system in the asymptotic regime. Consequently, the Lyapunov exponent can be written as
\begin{equation}
\lambda={\rm ln}(\rho). \label {expoente Lyapunov mapa linear}  
\end{equation}

\begin{figure}[hbt]
\centering
\includegraphics[scale=0.45]{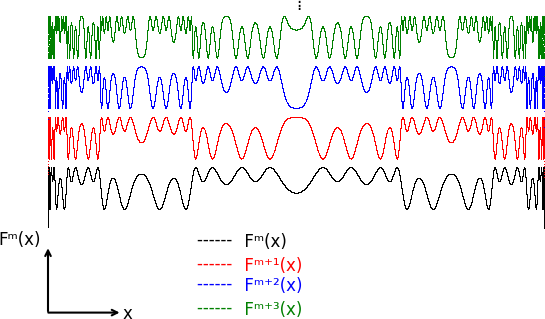}
\caption{Illustration of the mappings $F^{m}(x)$, $F^{m+1}(x)$, $F^{m+2}(x)$, and $F^{m+3}(x)$ of a chaotic system, which are related by the scale factor $\rho=e^{\lambda}$ in the asymptotic limit $m\rightarrow\infty$.}
\label{fig5}
\end{figure}

In the limit $m\rightarrow\infty$, the mappings of order $m$ form a geometric inverse set, as illustrated in Fig.~\ref{fig5}. These mappings behave analogously to those presented in Fig.~3 of Subsection~4.1.

According to Eq.~\eqref{comprimento2}, the lengths of the mappings of order $m\pm q$ are given by
\begin{equation}
L(m \pm q) \approx e^{(m \pm q)\lambda}\Delta x. \label {comprimentos mq}
\end{equation}

Combining Eqs.~\eqref{comprimento2} and \eqref{comprimentos mq}, we obtain
\begin{equation}
L(m+q).L(m-q)\approx L^2(m). \label {comprimentos assintoticos} 
\end{equation}

Equation~\eqref{comprimentos assintoticos} characterizes the inversion symmetry of the system in the asymptotic limit. The corresponding normalized inversion functions are
\begin{equation}
f_i^{(1)}(q)=\frac{L(m-q)}{L(m)}\approx e^{-q\lambda}, \label {funçoes normalizadas 1} 
\end{equation}
and
\begin{equation}
f_i^{(2)}(q)=-\frac{L(m+q)}{L(m)}\approx -e^{q\lambda}, \label {funçoes normalizadas 2} 
\end{equation}
which satisfy $f_i^{(1)}(q).f_i^{(2)}(q)=-1$.

In this limit, we consider the normalized inversion functions given in Eq.~\eqref{função de inversão com expoente} with $a_1=a_2=1$, yielding
\begin{equation}
f_i^{(1)}(\gamma)=e^{-\gamma}, \label {função de inversão limite assintotico 1} 
\end{equation}
and
\begin{equation}
f_i^{(2)}(\gamma)=-e^{\gamma}. \label {função de inversão limite assintotico 2} 
\end{equation}

By equating the length-based inversion functions in Eqs.~\eqref{funçoes normalizadas 1} and \eqref{funçoes normalizadas 2} with the normalized inversion functions in Eqs.~\eqref{função de inversão limite assintotico 1} and \eqref{função de inversão limite assintotico 2}, respectively, we obtain the relationship between the inversion exponent $\gamma$ and the Lyapunov exponent $\lambda$,
\begin{equation}
\gamma=q\lambda. \label {relação expoentes de inversao e Lyapunov} 
\end{equation}

Equation~\eqref{relação expoentes de inversao e Lyapunov} is valid for $\lambda\geq0$. When $\lambda=0$, the lengths of the mappings converge to a constant value, the scale factor tends to unity, and both the inversion exponent and the Lyapunov exponent vanish. This result can be inferred from Eq.~\eqref{comprimento} for $\lambda=0$, which yields the asymptotic length $L(m)\approx\sqrt{2}\,\Delta x$.

For negative Lyapunov exponents, Eq.~\eqref{comprimento} shows that $L(m)\approx\Delta x$, indicating that the lengths of the mappings again converge to a constant value and the inversion exponent vanishes. In this case, there is no direct relationship between the inversion and Lyapunov exponents, and Eq.~\eqref{relação expoentes de inversao e Lyapunov} is no longer valid.

Thus, in the asymptotic regime, the inversion exponent vanishes for periodic and quasiperiodic orbits. For chaotic orbits, however, the system exhibits scale symmetry, with the mappings asymptotically forming a geometric inverse set and the inversion exponent converging to the Lyapunov exponent.

To validate the proposed theoretical framework, we estimate the Lyapunov exponents using scale and inversion symmetries for selected parameter values of the tent map, the logistic map, and the Chebyshev map.

\subsubsection{Tent map}
The tent map \cite{Al:96,Ca:17} is defined as
\begin{center}
$F_T(x)=
\begin{cases}
2\nu x_n, & 0\le x_n\le \frac{1}{2}, \\
2\nu(1-x_n), & \frac{1}{2}<x_n\le 1,
\end{cases}$
\end{center}
with $0<\nu\le1$.
From Eq.~\eqref{expoente de Lyapunov}, the Lyapunov exponent is given by $\lambda={\rm ln}(2\nu)$, and the map exhibits chaotic behavior for $\nu>1/2$.

\begin{figure}[hbt]
\centering
\includegraphics[scale=0.45]{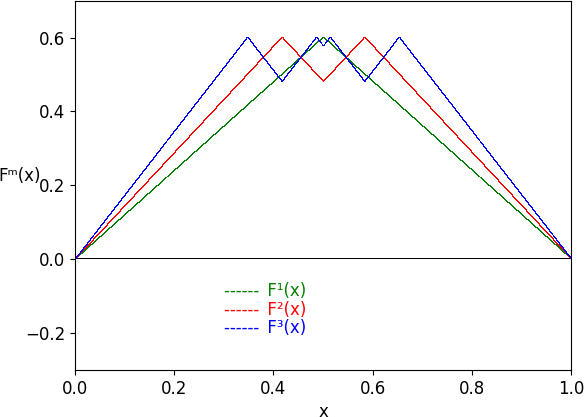}
\caption{Mappings $F^{1}(x)$, $F^{2}(x)$, and $F^{3}(x)$ for the tent map with parameter $\nu=0.6$.}
\label{fig6}
\end{figure}

Figure~\ref{fig6} shows the mappings $F^{m}(x)$ for $m=1,2,3$ of the tent map with parameter $\nu=0.6$.

To estimate the Lyapunov exponent of the tent map using scale and inversion symmetries, we consider the lengths of the mappings of order $m$. For the tent map, these lengths can be obtained analytically from
\begin{equation}
L(m)=\sqrt{1+\left(2\nu\right)^{2m}}, \quad m=1,2,\ldots
\end{equation}

In the asymptotic limit, Eq.~\eqref{comprimento} is recovered, yielding
\begin{equation}
L(m)\approx (2\nu)^m \approx e^{m\lambda}. \label {comprimento aproximado} 
\end{equation}
From Eq.~\eqref{comprimento aproximado}, the Lyapunov exponent is directly obtained as $\lambda={\rm ln}(2\nu)$.

In this regime, the scale symmetry of the mappings is expressed as
\begin{equation}
L(m+1)\approx (2\nu)L(m). \label {comprimento assintótico} 
\end{equation}

Equation~\eqref{comprimento assintótico} corresponds to a linear map with scale factor $\rho=2\nu$. Thus, the Lyapunov exponent follows from the scale symmetry via Eq.~\eqref{expoente Lyapunov mapa linear}, namely, $\lambda={\rm ln}(\rho)={\rm ln}(2\nu)$.

Inversion symmetry emerges when the relation between mapping lengths in the asymptotic regime satisfies Eq.~\eqref{comprimentos assintoticos}. For the tent map, the normalized inversion functions are given by
\begin{equation}
f_i^{(1)}(m,q)=\frac{\sqrt{1+(2\nu)^{2m-2q}}}{\sqrt{1+(2\nu)^{2m}}},
\end{equation}
and
\begin{equation}
f_i^{(2)}(m,q)=-\frac{\sqrt{1+(2\nu)^{2m+2q}}}{\sqrt{1+(2\nu)^{2m}}}.
\end{equation}

Considering the normalized inversion function $f_i^{(1)}(m,q)$ in the limit $m\rightarrow\infty$, we obtain
\[
\lim_{m\rightarrow\infty}f_i^{(1)}(m,q)=
\begin{cases}
1, & 0<\nu\le\frac{1}{2}, \\
(2\nu)^{-q}, & \frac{1}{2}<\nu\le1.
\end{cases}
\]

In this limit, the normalized inversion functions are given by Eqs.~\eqref{função de inversão limite assintotico 1} and \eqref{função de inversão limite assintotico 2}. Consequently, $e^{-\gamma}=(2\nu)^{-q}$, which leads to $\gamma=q\,{\rm ln}(2\nu)$.

\begin{figure}[hbt]
\centering
\includegraphics[scale=0.45]{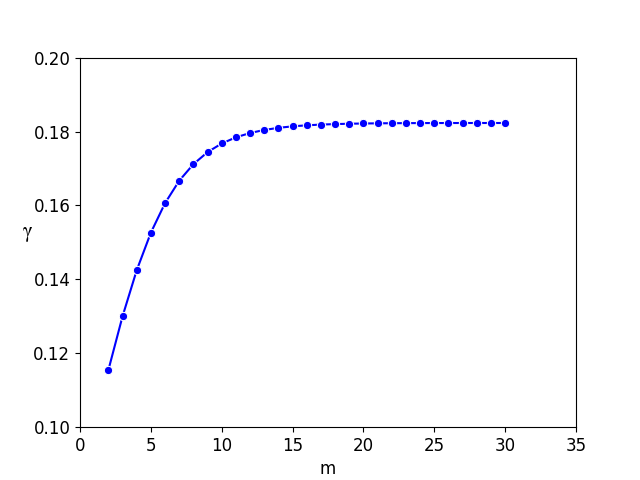}
\caption{Inversion exponent $\gamma$ as a function of $m$ for the tent map with parameter $\nu=0.6$.}
\label{fig7}
\end{figure}

Since $\lambda={\rm ln}(2\nu)$, the relation given in Eq.~\eqref{relação expoentes de inversao e Lyapunov} is satisfied, namely $\gamma=q\lambda$. Figure~\ref{fig7} illustrates the behavior of the inversion exponent $\gamma$ as a function of $m$ for the tent map with parameter $\nu=0.6$, considering the first 30 iterations. In the asymptotic regime, the scale factor converges to $\rho=1.2$, and $\gamma$ converges to $\lambda={\rm ln}(2\nu)\approx0.18$.
                                        
\begin{figure}[hbt]
\centering
\includegraphics[scale=0.45]{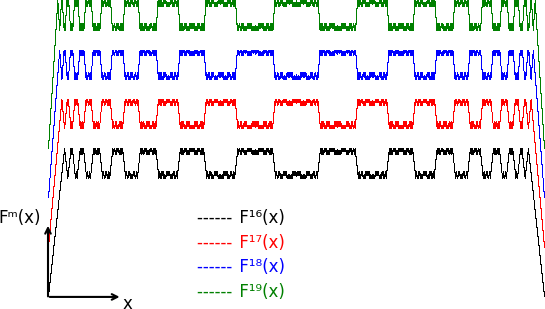}
\caption{Illustration of the mappings $F^{16}(x)$, $F^{17}(x)$, $F^{18}(x)$, and $F^{19}(x)$ for the tent map with parameter $\nu=0.6$.}
\label{fig8}
\end{figure}

Asymptotically, the mappings $F^{m}(x)$ belong to a geometric inverse set exhibiting scale and inversion symmetries. For $\nu=0.6$, this behavior can be observed for mappings of order $m=16$ to $m=19$, as shown in Fig.~\ref{fig8}, where the relationship between the mapping lengths is approximately governed by the scale factor $\rho=2\nu$.

For the parameter value $\nu=0.5$, we find that Eq.~\eqref{relação expoentes de inversao e Lyapunov} remains valid. In this case, the mapping length is constant and given by $L(m)=\sqrt{2}\Delta x=\sqrt{2}$, the inversion function is unitary, and both the inversion and Lyapunov exponents are zero.

For $\nu<0.5$, the Lyapunov exponent becomes negative. In this regime, the mapping lengths do not exhibit asymptotic scale symmetry, as they converge to a constant value, $L(m)\approx\Delta x\approx1$, and the inversion exponent vanishes.

\subsubsection{Logistic map}
The logistic map \cite{Al:96,Ca:17} is governed by
\begin{equation}
x_{n+1}=rx_n(1-x_n),
\end{equation}
where $0\le x_n\le1$ and $r$ is the control parameter, with $0\le r\le4$. The Lyapunov exponent of the logistic map is computed directly from Eq.~\eqref{expoente de Lyapunov}.

As observed in the case of the tent map, in the asymptotic limit we can determine the Lyapunov exponent in the chaotic regime by exploiting scale and inversion symmetries using the mappings $F^{m}(x)$ of order $m$.

\begin{figure}[hbt]
\centering
\includegraphics[scale=0.45]{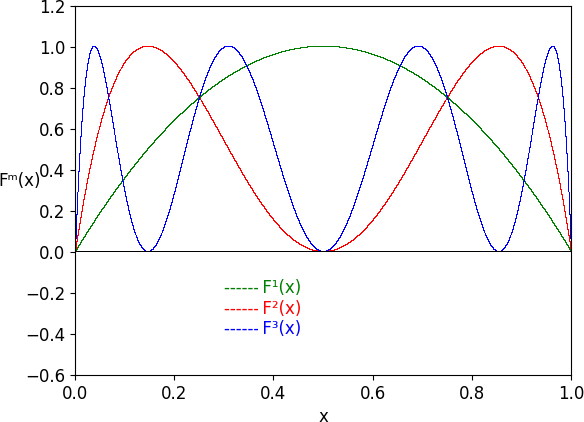}
\caption{Mappings $F^{1}(x)$, $F^{2}(x)$, and $F^{3}(x)$ for the logistic map with parameter $r=4$.}
\label{fig9}
\end{figure}

Figure~\ref{fig9} shows the mappings $F^{m}(x)$ for $m=1,2,3$ with $r=4$, as an illustrative example.

To estimate the inversion exponents, we computed the lengths of the mappings $F^{m}(x)$ for orders $m=1$ to $m=17$, considering an interval $\sigma_x=6.67\times10^{-9}$ between consecutive points along each mapping.

\begin{figure}[hbt]
\centering
\includegraphics[scale=0.45]{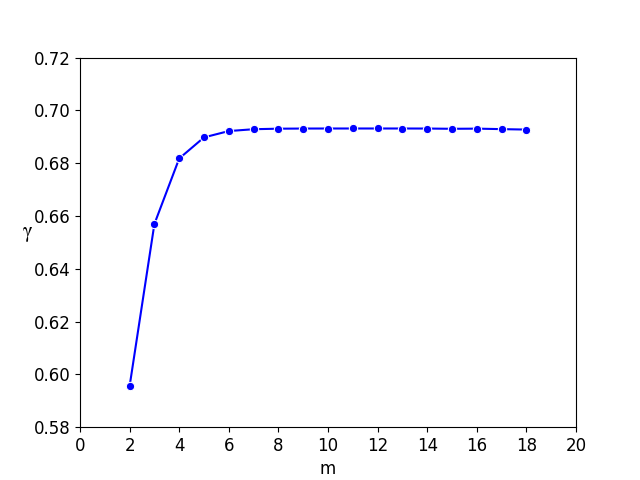}
\caption{Variation of the inversion exponent $\gamma$ as a function of the iteration $m$ for the logistic map with parameter $r=4$.}
\label{fig10}
\end{figure}

Figure~\ref{fig10} shows the evolution of the inversion exponent for $r=4$. We observe that from $m=4$ onward the inversion exponent converges to $\lambda=\ln(2)\approx0.69$. These results indicate that the proposed method offers a clear advantage over the standard approach, as convergence is achieved with very few iterations. Nevertheless, in situations involving more complex mappings or requiring higher numerical precision, a larger number of discretization points may be needed to estimate the curve lengths accurately, leading to increased computational cost.

\begin{figure}[hbt]
\centering
\includegraphics[scale=0.45]{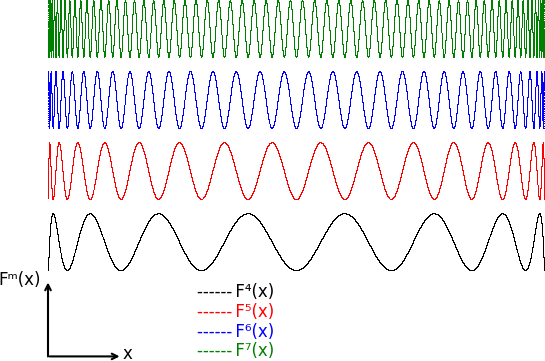}
\caption{Illustration of the mappings $F^{4}(x)$, $F^{5}(x)$, $F^{6}(x)$, and $F^{7}(x)$ for the logistic map with parameter $r=4$.}
\label{fig11}
\end{figure}

The geometric inverse set associated with the logistic map for $r=4$ is characterized by the scale factor $\rho=e^{\lambda}=2$. Figure~\ref{fig11} illustrates the mappings $F^{4}(x)$ to $F^{7}(x)$ belonging to this inverse set.

For parameter values yielding Lyapunov exponents $\lambda\leq0$, we may consider, for instance, the cases $r=1$ and $r=0.5$, for which the Lyapunov exponents are $\lambda=0$ and $\lambda\approx-0.69$, respectively. Since the Lyapunov exponent of the logistic map is obtained as a numerically averaged quantity, in both cases the corresponding mapping lengths tend to $L(m)\approx\Delta x\approx1$, and the inversion exponent vanishes.

\subsubsection{Chebyshev map}
The Chebyshev map \cite{Ge:84,Zh:24} is defined as
\begin{equation}
x_{n+1}={\rm cos}\left(\beta\,{\rm cos}^{-1}(x_n)\right),
\end{equation}
where $\beta\in\mathbb{N}$ and $-1\le x_n\le1$. The Lyapunov exponent of this map is $\lambda={\rm ln}(\beta)$.

\begin{figure}[hbt]
\centering
\includegraphics[scale=0.45]{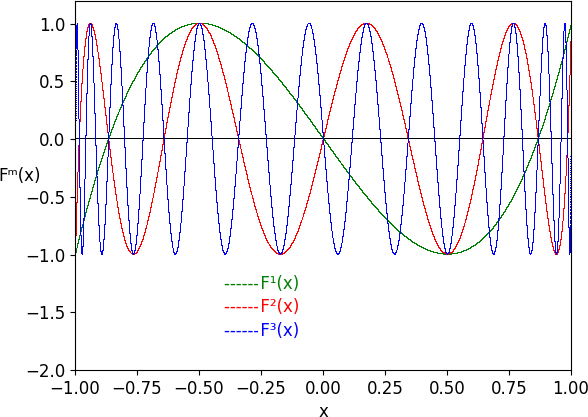}
\caption{Mappings $F^{1}(x)$, $F^{2}(x)$, and $F^{3}(x)$ for the Chebyshev map with parameter $\beta=3$.}
\label{fig12}
\end{figure}

Figure~\ref{fig12} shows the mappings $F^{m}(x)$ for $m=1,2,3$ with $\beta=3$, as an illustrative example.

To obtain the inversion exponents, we computed the lengths of the mappings $F^{m}(x)$ for orders $m=1$ to $m=10$, considering an interval $\sigma_x=10^{-8}$ between consecutive points along each mapping.

\begin{figure}[hbt]
\centering
\includegraphics[scale=0.45]{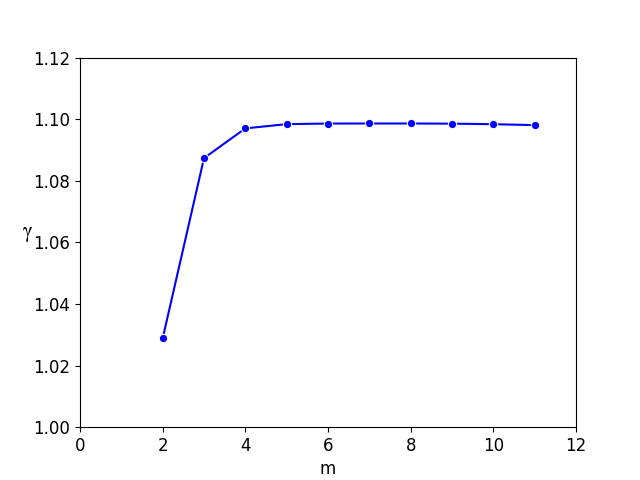}
\caption{Variation of the inversion exponent $\gamma$ as a function of the iteration $m$ for the Chebyshev map with parameter $\beta=3$.}
\label{fig13}
\end{figure}

Figure~\ref{fig13} shows the inversion exponent for the first ten iterations with $\beta=3$. We observe that the inversion exponent converges asymptotically to $\lambda={\rm ln}(3)\approx1.10$.

\begin{figure}[hbt]
\centering
\includegraphics[scale=0.45]{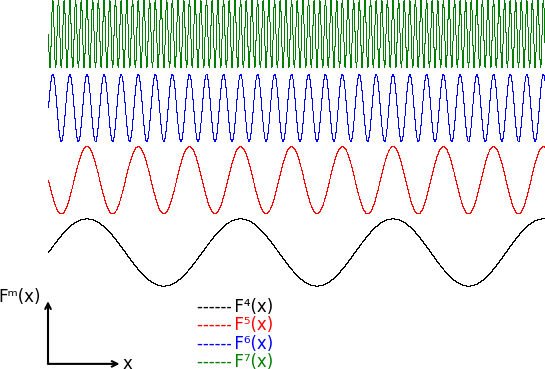}
\caption{Illustration of the mappings $F^{4}(x)$, $F^{5}(x)$, $F^{6}(x)$, and $F^{7}(x)$ for the Chebyshev map with parameter $\beta=3$, where the interval $\Delta x=0.25$ is used for improved graphical visualization.}
\label{fig14}
\end{figure}

The geometric inverse set associated with the Chebyshev map for $\beta=3$ is characterized by a scale factor $\rho=e^{\lambda}=3$. Figure~\ref{fig14} displays the mappings $F^{4}(x)$ to $F^{7}(x)$ belonging to this inverse set.

The Lyapunov exponent of the Chebyshev map vanishes only for $\beta=1$. In this case, the mapping lengths remain constant for all iterations and are given by $L(m)=\sqrt{2}\Delta x=2\sqrt{2}$. Consequently, the inversion exponent is zero, in agreement with the Lyapunov exponent.

For negative Lyapunov exponents, we may consider, for example, the case $\beta=0.1$, for which $\lambda\approx-2.03$. In this regime, the mapping lengths converge to $L(m)\approx\Delta x\approx2$, and the inversion exponent is again $\gamma=0$.

\section{Conclusions}\label{sec_conclusion}
In this work, we investigated scale invariance in discrete dynamical systems and fractals by exploring the interplay between scale and inversion transformations. To analyze inversion symmetry, we extended the notion of geometric inversion and introduced the concepts of inverse sets, inversion resultants, inversion functions, and inversion differential equations. These definitions were then employed to characterize inversion symmetry in self-similar fractals and one-dimensional chaotic maps.

We first examined inverse sets generated by geometric figures and identified the differential equations associated with the inversion symmetry of these figures, as well as of self-similar fractals. In the case of self-similar fractals, we demonstrated that, in addition to scale symmetry, these systems also possess inversion symmetry. Measurements of length, area, and volume in self-similar fractals can be described in terms of inversion symmetry, leading to differential equations whose solutions are exponential laws expressed as functions of the fractal dimension. Consequently, we showed that the scale invariance of fractals can be formulated both through power laws, when scale symmetry is considered alone, and through exponential laws, when inversion symmetry is taken into account.

For discrete dynamical systems, we showed that the mappings of order $m$ associated with one-dimensional chaotic maps can exhibit scale and inversion symmetries in the asymptotic limit. This property enables the computation of Lyapunov exponents through scale and inversion symmetries. In this regime, we established a direct relationship between the inversion exponent $\gamma$ and the Lyapunov exponent $\lambda$ for positive and vanishing Lyapunov exponents. We verified that, asymptotically, the inversion exponent vanishes for periodic and quasiperiodic orbits. For chaotic systems, however, the inversion exponent converges to the Lyapunov exponent, and the $m$-order mappings evolve within a geometric set characterized by scale and inversion symmetries. In this sense, chaotic systems may be described not only by strange attractors with fractal geometry, but also by geometric sets exhibiting asymptotic scale and inversion symmetries.

Using the proposed method to compute positive Lyapunov exponents, we observed a rapid convergence of the inversion exponent toward the Lyapunov exponent for the maps analyzed in this work. Moreover, the method provides a standardized and conceptually transparent procedure for estimating Lyapunov exponents, requiring only the calculation of mapping lengths and the verification of their asymptotic scale and inversion symmetries. This represents a practical advantage over traditional approaches based on long-time averages of local derivatives.

Overall, our results support the view that scale invariance is an intrinsic feature of discrete dynamical systems exhibiting chaotic behavior, and that positive Lyapunov exponents can be efficiently estimated through the combined use of scale and inversion symmetries.

\section*{Acknowledgements}
The authors acknowledge financial support from the Brazilian Federal Agencies Conselho Nacional de Desenvolvimento Cient\'ifico e Tecnol\'ogico (CNPq), grant No.~302665/2017-0; the S\~ao Paulo Research Foundation (FAPESP), grant No.~2024/05700-5; and the Coordena\c{c}\~ao de Aperfei\c{c}oamento de Pessoal de N\'ivel Superior (CAPES). E.D.L.~acknowledges support from CNPq (Nos.~301318/2019-0 and 304398/2023-3) and FAPESP (Nos.~2019/14038-6 and 2021/09519-5). E.C.G.~acknowledges financial support from FAPESP under grant No.~2025/02318-5.

\section*{DATA AVAILABILITY}
The data that support the findings of this study are available from the corresponding author upon reasonable request.


\end{document}